\documentclass[sigconf,nonacm]{acmart}
\AtBeginDocument{%
  \providecommand\BibTeX{{%
    \normalfont B\kern-0.5em{\scshape i\kern-0.25em b}\kern-0.8em\TeX}}}

\setcopyright{acmlicensed}
\copyrightyear{2026}
\acmYear{2026}
\acmDOI{XXXXXXX.XXXXXXX}

\acmConference[MSR 2026]{23rd International Mining Software Repositories Conference}{April 13--14,2026}{Rio de Janeiro, Brazil}

\acmISBN{978-1-4503-XXXX-X/18/06}

\usepackage{tcolorbox}
\usepackage{xcolor}
\usepackage{graphicx}
\usepackage{textcomp}
\usepackage{braket}
\usepackage{physics}
\usepackage{listings}
\usepackage{siunitx}
\usepackage{booktabs}
\usepackage{subcaption}
\usepackage[noend]{algpseudocode}
\usepackage{algorithm}
\usepackage{hyperref}
\usepackage{url}
\usepackage{xcolor}
\usepackage{color}
\usepackage{xspace}
\usepackage{multicol}
\usepackage{tabularx}
\usepackage{cprotect}
\usepackage{graphicx}
\usepackage{lscape}
\usepackage{rotating}
\usepackage{seqsplit}

\usepackage{booktabs}   
\usepackage{xurl}     
\usepackage[utf8]{inputenc}
\usepackage{longtable}
\usepackage{array}

\usepackage{multirow}
\usepackage[T1]{fontenc}
\usepackage{lmodern}
\usepackage{times}

\usepackage{minted} 
\usepackage{geometry}
\usepackage{ragged2e}

\newtcolorbox{researchbox}{
  colback=blue!5!white,
  colframe=blue!75!black,
  fonttitle=\bfseries,
  title=Research Question,
  width=\linewidth,
  arc=2mm,
  boxrule=0.5pt
}

\setminted{
    bgcolor=black!5, 
    fontsize=\footnotesize,
    linenos,         
    frame=lines,     
    breaklines,      
    escapeinside=||  
}

\hypersetup{
    colorlinks=true,
    linkcolor=blue,
    filecolor=magenta,      
    urlcolor=cyan,
}

\usepackage{breakurl}

\definecolor{backcolor}{rgb}{0.95,0.95,0.92}
\definecolor{codegray}{rgb}{0.5,0.5,0.5}
\definecolor{codegreen}{rgb}{0,0.6,0}
\definecolor{codeblue}{rgb}{0.015,0.015,0.66}
\definecolor{codepurple}{rgb}{0.58,0,0.82}
\definecolor{codered}{rgb}{1.0, 0.0, 0}

\lstdefinestyle{mystyle}{
    backgroundcolor=\color{black!5},
    commentstyle=\color{green!40!black},
    keywordstyle=\color{blue},
    numberstyle=\tiny\color{gray},
    stringstyle=\color{purple},
    basicstyle=\ttfamily\footnotesize,
    breakatwhitespace=false,
    breaklines=true,
    captionpos=b,
    keepspaces=true,
    numbers=left,
    numbersep=5pt,
    showspaces=false,
    showstringspaces=false,
    showtabs=false,
    tabsize=2
}
\begin{document}

\title{Mining Quantum Software Patterns in Open-Source Projects}

\author{Neilson C. L. Ramalho}
\email{neilson@usp.br}
\affiliation{%
  \institution{School of Arts, Sciences, and Humanities, University of São Paulo}
  \country{Brazil}
}

\author{Erico A. da Silva}
\email{augusto.ericosilva@usp.br}
\affiliation{%
  \institution{School of Arts, Sciences, and Humanities, University of São Paulo}
  \country{Brazil}
}

\author{Higor Amario de Souza}
\email{higoramario@usp.br}
\affiliation{%
  \institution{Department of Computing and Digital Systems Engineering, Polytechnic School, University of São Paulo}
  \country{Brazil}
}

\author{Marcos Lordello Chaim}
\email{chaim@usp.br}
\affiliation{%
  \institution{School of Arts, Sciences, and Humanities, University of São Paulo}
  \country{Brazil}
}

\begin{abstract}
Quantum computing has become an active research field in recent years, as its applications in fields such as cryptography, optimization, and materials science are promising. Along with these developments, challenges and opportunities exist in the field of Quantum Software Engineering, as the development of frameworks and higher-level abstractions has attracted practitioners from diverse backgrounds. Unlike initial quantum frameworks based on the circuit model, recent frameworks and libraries leverage higher-level abstractions for creating quantum programs. 
This paper presents an empirical study of 985 Jupyter Notebooks from 80 open-source projects to investigate how quantum patterns are applied in practice. Our work involved two main stages. First, we built a knowledge base from three quantum computing frameworks (Qiskit, PennyLane, and Classiq). This process led us to identify and document 9 new patterns that refine and extend the existing quantum computing pattern catalog. Second, we developed a reusable semantic search tool to automatically detect these patterns across our large-scale dataset, providing a practitioner-focused analysis.
Our results show that developers use patterns in three levels: from foundational circuit utilities, to common algorithmic primitives (e.g., Amplitude Amplification), up to domain-specific applications for finance and optimization. This indicates a maturing field where developers are increasingly using high-level building blocks to solve real-world problems.
\end{abstract}



\maketitle

\section{Introduction}
Quantum computing (QC) has recently left the pure theoretical realm to become an active research field, with heavy investments from major companies. As technology evolves, especially in regard to hardware and noise reduction, practitioners have seen the possibility to use Quantum Computing's capabilities to solve problems previously considered unmanageable for classical computers in areas like molecular simulations, cybersecurity, finance, and logistics. The main concept behind a quantum computer is to take advantage of specific quantum mechanics properties to perform computation \cite{Hidary2019Quantum}.

Although building quantum algorithms is a promising area, developing these algorithms based on circuits poses different challenges for engineers not familiar with quantum mechanics concepts such as superposition, entanglement, interference, and the probabilistic nature of quantum computers. 
Classical computing has evolved over the years, providing engineers with a mature ecosystem of tools and patterns. However, the quantum domain currently lacks the development procedures and standards prevalent in classical computing, a gap that the field of Quantum Software Engineering aims to fill \cite{MurilloQSE2025}.
In order to move away from the circuit-based computing model, several authors have studied higher-level abstractions for creating quantum algorithms. These higher-level abstractions manifest themselves as quantum computing patterns \cite{LeymannQuantumAlgorithms, PatternsCircuitCutting2023, HarnessingPatterns2025, PatternsQuantumErrorHandling2022, Georg2023_PatternsQuantumExecution, Buehler2023_QuantumSoftwareEngineeringPatterns, Stiliadou2025_QMLPatterns}.

This paper examines the patterns that practitioners use in open-source quantum projects. More specifically, our investigation aims at finding which high-level QC patterns are being used by practitioners across different repositories of open-source projects that use
QC Python-based frameworks. 
To guide our analysis, we formulated the following research questions.

\begin{itemize}
    \item [\textbf{RQ1.}] What are the most common QC concepts used in popular open-source frameworks?
    \item [\textbf{RQ2.}] What QC patterns appear in practice that are not yet documented in the literature?
    \item [\textbf{RQ3.}] What QC architectural patterns are observed in these frameworks?
    \item [\textbf{RQ4.}] How are QC patterns applied to develop practical applications, as opposed to internal framework code?
\end{itemize}

We divided our work into two main stages. First, we needed to connect the abstract pattern descriptions to actual code. To do so, we built a catalog of high-level components by inspecting the source code of three popular frameworks: Classiq SDK (version 0.88.0) \cite{classiq-library2024}, PennyLane (version 0.44.0)\cite{pennylane_qml_github}, and Qiskit (version 2.3.0)\cite{qiskit2024}. We then categorized each component against the patterns documented on the Quantum Pattern Atlas website \cite{PlanQK_QuantumPatterns_2024}. We found that the existing list of patterns was incomplete in classifying all the components we identified. Therefore, we define and add nine new patterns.
The result of this first stage is a knowledge base that links patterns to concrete code examples. This database can be used as a reference by researchers, practitioners, and students.

Second, we used this new, enriched catalog to analyze a list of 985 Jupyter Notebooks from 80 different open-source projects. We developed a tool that uses embedding models based on sentence transformers \cite{reimers-2019-sentence-bert} to automatically search these projects for our listed patterns. The goal of this search was to determine which patterns are used most frequently in practice.
Our analysis found that only 24 of the 61 catalog patterns were present in the QC frameworks. Of these, 22 occur in the Jupyter Notebooks, including all nine new QC patterns. This finding suggests many QC patterns do not materialize in QC applications. It also shows that practitioners are adopting higher-level abstractions at three different levels. At a foundational level, we observed heavy reliance on circuit construction utilities, indicating that fine-grained control remains important. Above this, the most frequently used patterns are algorithmic primitives, such as the Quantum Fourier Transform (QFT) and Quantum Phase Estimation (QPE), which serve as reusable building blocks. Finally, at the highest level, we find domain-specific application patterns for optimization and finance, suggesting a growing focus on solving practical problems. This layered adoption indicates a maturing ecosystem where developers are increasingly moving from low-level gate manipulation to problem-solving with high-level components.

By mining open-source quantum repositories, we achieved the following contributions. (1) A knowledge base with QC patterns mapped to concrete implementations that can be used as a reference for researchers and students. (2) Nine new QC patterns identified by inspecting popular QC frameworks and grouping existing patterns into composite structures. (3) A reusable semantic tool developed to find the implementations and patterns of quantum algorithms in Python-based QC projects. (4) Insights on how QC patterns occur in 985 Jupyter Notebooks from 80 open-source projects.

This work is organized as follows. Section \ref{sec:background} provides a short background and an overview of the current research on Quantum Software Patterns. 
Section \ref{sec:experiment} details the mining process we conducted to help answer the defined research questions. The results and discussion are presented in Section \ref{sec:results}. 
The related work is presented in Section \ref{related_work}. We conclude the paper with possible research directions in Section \ref{sec:conclusions}.

\section{Background}
\label{sec:background}
A quantum computer is a device that leverages the properties of quantum mechanics to perform computation \cite{Hidary2019Quantum}. Properties such as entanglement, superposition and interference are useful in a class of problems that are intractable by classical computers.
The intersection between quantum mechanics and computation brings new challenges to classical software engineering practices, especially in understanding and handling QC properties and the probabilistic nature of quantum computers. While classical software engineering has built a robust theoretical and practical basis over recent decades, such practices for quantum computing are still emerging. To bridge this gap, the research community in Quantum Software Engineering is creating higher-level abstractions and defining reusable patterns \cite{LeymannQuantumAlgorithms, PatternsCircuitCutting2023, HarnessingPatterns2025, PatternsQuantumErrorHandling2022, Georg2023_PatternsQuantumExecution, Buehler2023_QuantumSoftwareEngineeringPatterns, Stiliadou2025_QMLPatterns}. Practitioners are now using frameworks that leverage these patterns and provide abstractions to implement hybrid quantum programs (QPs) for real-world problems. 

This section presents an overview of the recent research in Quantum Computing Patterns.

\subsection{Quantum Computing Software Patterns}

Several authors have explored software patterns for quantum computing. For instance, patterns have been proposed for core quantum operations, such as Initialization to prepare a quantum register, Uniform Superposition to create an equal superposition of all basis states, and Amplitude Amplification to increase the measurement probability of a desired state \cite{LeymannQuantumAlgorithms}. In addition to algorithm construction, research also focuses on the physical limitations of quantum hardware. This has led to patterns for error management, which include methods for Error Correction using redundancy, as well as mitigation strategies for errors that occur during gate execution or final measurement \cite{PatternsQuantumErrorHandling2022}.  Other work contributes patterns for circuit execution strategies for running quantum applications on cloud services \cite{Georg2023_PatternsQuantumExecution}. These patterns address scenarios ranging from the Standalone Circuit Execution for single circuits to the Orchestrated Execution for managing hybrid applications using workflows. There are also patterns for quantum circuit cutting, an approach that divides a quantum computation into multiple smaller parts to execute on devices with limited qubits. This work includes the definition of a Circuit Cutting pattern, which is refined by specific techniques such as the Gate Cut, for decomposing a multi-qubit gate, and the Wire Cut, for interrupting a wire between gates \cite{PatternsCircuitCutting2023}.
Building on these concepts, other researchers have defined patterns for higher-level software engineering and architectural concerns. To promote reusability and manage hybrid systems, patterns like the Quantum Module encapsulate a quantum algorithm into a reusable component, while the Hybrid Module packages quantum and classical parts into an integrated unit \cite{Buehler2023_QuantumSoftwareEngineeringPatterns}. At the system architecture level, patterns address the integration of quantum components within larger classical systems. Examples include the Quantum-Classical Split and the Quantum Middleware Layer, which facilitates interaction between quantum and classical resources in Quantum AI systems \cite{klymenko2024architecturalpatternsdesigningquantum}. Finally, patterns have also been developed for specific application domains, such as Quantum Machine Learning (QML), covering areas like Quantum Clustering, Classification, and Neural Networks \cite{Stiliadou2025_QMLPatterns}. These and other patterns are summarized in Table~\ref{tab:survey_of_patterns}.

\begin{table*}[t] %
    \centering
    \caption{Summary of Quantum Software Patterns}
    \label{tab:survey_of_patterns}
    \begin{tabularx}{\linewidth}{@{} l l >{\raggedright\arraybackslash}X @{}}
    \toprule
    \textbf{Category / Source} & \textbf{Pattern Name} & \textbf{Brief Description} \\
    \midrule
    \multicolumn{3}{@{}l}{\textit{Foundational Algorithm Patterns \cite{LeymannQuantumAlgorithms}}} \\
     & Initialization & Prepares the initial state of a quantum register \\
     & Uniform Superposition & Creates an equal superposition of all basis states \\
     & Oracle & A ``black-box'' operation that marks solution states \\
     & Amplitude Amplification & Boosts the measurement probability of a desired state \\
     & \multicolumn{2}{l}{... (and others: Creating Entanglement, Function Table, etc.)} \\
    \cmidrule(r){1-3}
    \multicolumn{3}{@{}l}{\textit{Error Management Patterns \cite{PatternsQuantumErrorHandling2022}}} \\
     & Error Correction & Detects and corrects errors during computation using redundancy. \\
     & Readout Error Mitigation & Reduces errors that occur during the final measurement \\
     & Gate Error Mitigation & Mitigates errors from imperfect or noisy gate execution \\
    \cmidrule(r){1-3}
    \multicolumn{3}{@{}l}{\textit{Hybrid Solutions \& Reusability Patterns \cite{Buehler2023_QuantumSoftwareEngineeringPatterns}}} \\
     & Quantum Module & Encapsulates a quantum algorithm into a reusable component. \\
     & Hybrid Module & Packages quantum and classical parts into an integrated module \\
     & Quantum Circuit Translation & Translates a circuit to enable interoperability \\
    \cmidrule(r){1-3}
    \multicolumn{3}{@{}l}{\textit{Quantum AI Architectural Patterns \cite{klymenko2024architecturalpatternsdesigningquantum}}} \\
     & Quantum-Classical Split & Integrates quantum components within classical systems \\
     & Quantum Middleware Layer & Facilitates interaction between quantum and classical resources \\
    \bottomrule
    \end{tabularx}
\end{table*}

\subsection{Quantum Computing Patterns Atlas}
\label{sub:pattern_atlas}
The list of patterns used in this work is based on the Quantum Computing Patterns Atlas, which documents 59 quantum computing patterns \cite{PlanQK_QuantumPatterns_2024}. Similar to the work by \citeauthor{DesignPatternsGof} \cite{DesignPatternsGof}, each pattern is cataloged with its intent, context, forces, solution, and related patterns, among other attributes.
The patterns cover multiple levels of abstraction. Some focus on basic circuit-level operations and techniques, such as Initialization, Uniform Superposition, Uncompute, and Circuit Cutting. Others work as reusable algorithmic building blocks that form the core of larger quantum algorithms, including Quantum Fourier Transformation, Amplitude Amplification, and Oracle. There are also patterns that cover higher-level application architectures and execution strategies for hybrid quantum-classical systems, such as Quantum Hardware Selection, Readout Error Mitigation, and Orchestrated Execution.




\section{Mining QC patterns}
\label{sec:experiment}
Our mining process aims to find implementations of the quantum patterns described in Section \ref{sub:pattern_atlas} within Python-based, open-source repositories. 
To bridge the gap between quantum concepts and their practical implementations, we searched for frameworks that offer a well-defined, centralized library of high-level, reusable components. Our selection criterion was that a framework must provide code excerpts (functions, classes or methods) dedicated to concepts as state preparations, ansätzes, or algorithmic building blocks, rather than focusing purely on gate-level circuit construction. These code excerpts are candidates to be classified as quantum patterns.

Based on this criterion, we selected three frameworks, PennyLane, Qiskit, and the Classiq Software Development Kit (SDK), which 
provide collections of high-level components. 
Furthermore, these frameworks are representative of the ecosystem. Qiskit, for instance, is the most popular QC framework on GitHub by star count. The number of stars that a repository has indicates how many users have marked that repository as a favorite. As for PennyLane, it is a reference in terms of QML algorithms and contains an extensive set of reusable classes that implement common QML routines. Similarly, Classiq SDK was selected because its library contains the most implementations of algorithms referenced on the Quantum Algorithm Zoo catalog \cite{QuantumZoo} (19 implementations in total as of September 2025).
In contrast, other frameworks such as Cirq were not included in this stage of the analysis. While popular (in terms of stars), Cirq does not offer a centralized library of pre-packaged, high-level abstractions.

\begin{figure}[h!]
    \centering
    \includegraphics[width=0.5\textwidth]{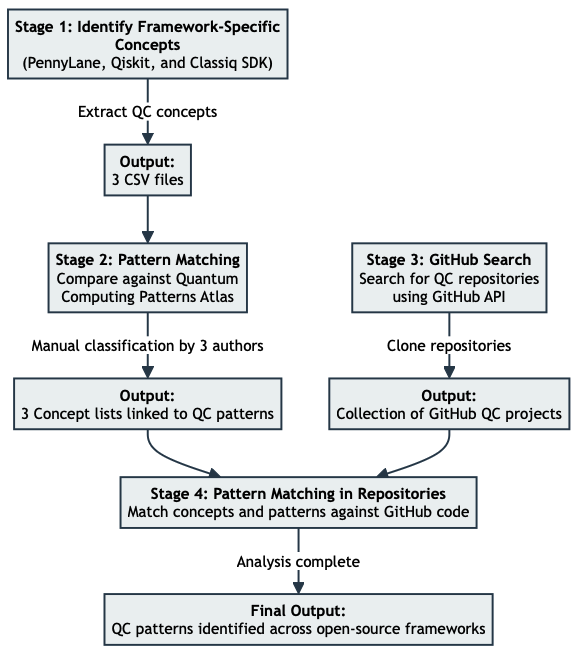}
    \caption{Overview of the 
    mining process.
    It consists of four main stages: (1) Identifying QC concepts in selected frameworks, (2) Matching those concepts to a known atlas of patterns, (3) Searching GitHub for a broad set of QC projects, and (4) Matching the identified patterns against the code from the GitHub repositories.}
    \label{fig:experiment_flow}
\end{figure}

By identifying how these three quantum computing frameworks implement quantum patterns, we created a tool that uses these implementations as references to search for similar concepts in a wider range of open-source quantum computing frameworks.
Figure \ref{fig:experiment_flow} provides an overview of our four-stage mining process.
\begin{enumerate}
    \item \textbf{Identify framework-specific concepts.}  We analyzed the source code of PennyLane, Qiskit, and Classiq SDK to identify how each framework implements and organizes high-level libraries and reusable components. For each framework, we created an automated process to extract and compile a list of implemented QC concepts. This produced a list of reusable quantum components (classes, methods, and functions) for each framework. We then exported the three lists as CSV files. Each file contained two columns: the full path of the concept (method, class, or function) and its description from the concept's docstring. A docstring is an embedded documentation string that describes a Python function, class, or module.

    \item \textbf{Pattern matching and consolidation:} In this stage, we compared each list of quantum concepts from stage one against a master list from the Quantum Computing Patterns Atlas (Section \ref{sub:pattern_atlas}). This matching process was carried out manually to determine which quantum patterns were present in each framework. 
    The first three authors of this work independently classified each QC concept. Discrepancies were reviewed
    by them 
    to reach a consensus. When no match was found, they discussed and defined new categories collectively. The result of this step was three lists of QC concepts, each linked to its corresponding pattern.
    
    \item \textbf{GitHub search for QC projects and repositories.} We used GitHub's API to search for QC repositories. We then used a script to clone the code from the found repositories.

    \item \textbf{QC pattern matching in GitHub repositories.} We created a tool that matches the lists of QC concepts and their associated QC patterns from the previous stages against the code in the GitHub repositories.
\end{enumerate}

The following sections detail each stage.

\subsection{Identify framework-specific concepts}
\label{sub:identifyQuantumConcepts}
This section describes how we identified high-level, reusable quantum components in PennyLane, Qiskit, and the Classiq SDK. Our goal was to create a catalog of the building blocks implemented in each framework. We searched for components that allow developers to write quantum programs without directly managing gate-level operations. Since the code organization and conventions of each framework differ, we programmatically analyzed each one with a custom approach. The specific methods and findings for each framework are detailed in the following subsections.

\subsubsection{Classiq}
\label{classiq_core_concepts_identification}
For the Classiq framework, our analysis targets the reusable components within its SDK, which is available on the Python Package Index (PyPI). This is distinct from Classiq's public GitHub library, which contains complete implementations of quantum algorithms. The focus of this work is not on the end-to-end algorithms, but on the building blocks used to construct them. The public API for these components is defined in the \_\_all\_\_ variable within the \_\_init\_\_.py file in the \textit{classiq/open\_library/functions/} package. To create a catalog with these components, a Python script was developed to dynamically import this public API list and then parse the relevant SDK source files to extract the definition and docstring for each function.


As of the writing of this paper, our process identified 60 quantum components, which are detailed in the supplementary material\footnote{The data supporting this study are publicly available in the \textit{quantum\_patterns} repository on  \url{https://anonymous.4open.science/r/quantum_patterns/}.}.


\subsubsection{PennyLane}
\label{pennylane}
For the PennyLane framework, the analysis focused on its \texttt{pennylane.templates} module. This module is the library's location for high-level, reusable circuit patterns such as state preparations, ansätzes, and embeddings. The methodology for identifying these components consisted of programmatically parsing the source code in this specific directory. A Python script was developed to scan for all Python classes that contain a docstring, as these represent the public, documented building blocks of the templates library. For each identified class, the script extracted its fully qualified name and complete docstring to create the component catalog.
The script's logic considered any documented class in the \texttt{pennylane.templates} folder as a reusable component. This process identified 68 quantum components from the PennyLane framework, which are detailed in the supplemental material.

\subsubsection{Qiskit}
\label{qiskit}
For the Qiskit framework, our analysis focuses on the \texttt{qiskit.circuit.library} module, which contains a collection of pre-built circuit components. To create a catalog of high-level, reusable components, a Python script was developed to programmatically parse the source code and filter out the non-relevant folders and files. The first step is to exclude two subdirectories: \texttt{standard\_gates}, which contains implementations for basic quantum gates, and templates, which consists of specific, non-general circuit patterns. The second step is to search reusable components in the remaining directories by extracting all public classes and standalone functions that contain a docstring, while ignoring private, underscore-prefixed functions. 

The last step was a two-stage de-duplication process. First, we removed redundant concepts represented by functions and classes with the same docstring and similar name structure. 
Subsequently, a semantic filtering stage was applied using the \texttt{all-mpnet-base-v2} Sentence Transformer model \cite{allmpnetbasev2} to group concepts with highly similar summary descriptions into groups. Within each cluster, a single canonical concept was preserved based on a heuristic that prioritizes non-deprecated classes (such as Gate subclasses) over functional aliases or duplicate implementations. 
This process identified 86 quantum components from the QisKit framework, which are also detailed in the supplemental material.

\subsection{Pattern matching and consolidation}
\label{sub:pattern_matching}
This step consisted of manually classifying the quantum concepts from the target frameworks against the pattern catalog described in Section \ref{sub:pattern_atlas}. The CSV files with the pattern catalog for each framework were converted to spreadsheets and shared with three authors. Each author then assigned a Quantum Pattern to each quantum concept from each individual catalog (PennyLane, Qiskit, and Classiq).
For each catalog, the classifications were 
compared and discussed, especially for the cases where there was no consensus\footnote{The resulting knowledge base can be verified on the CSV file available on \url{https://anonymous.4open.science/r/quantum_patterns/data/knowledge_base/knowledge_base.csv}}.



\begin{table}[h!]
\centering
\caption{Summary of the top 10 Quantum Patterns and their respective concept counts across Classiq, PennyLane, and Qiskit.}
\label{tab:combined_quantum_patterns}
\begin{tabularx}{\columnwidth}{X c}
\toprule
\textbf{Quantum Pattern} & \textbf{\begin{tabular}{@{}c@{}}Number of\\Concepts\end{tabular}} \\
\midrule
\multicolumn{2}{c}{\textbf{Classiq}} \\
\midrule
Basis Change & 8 \\
Quantum Arithmetic & 6 \\
Data Encoding & 6 \\
Amplitude Amplification & 5 \\
QSVT & 5 \\
Initialization & 5 \\
QAOA & 4 \\
Linear Combination of Unitaries & 2 \\
Circuit Construction Utility & 2 \\
QPE & 2 \\
\midrule
\textbf{Total} & \textbf{45} \\
\midrule
\multicolumn{2}{c}{\textbf{PennyLane}} \\
\midrule
Data Encoding & 9 \\
Quantum Arithmetic & 9 \\
Variational Quantum Eigensolver (VQE) & 7 \\
Variational Quantum Algorithm (VQA) & 5 \\
Quantum Neural Network (QNN) & 5 \\
Hamiltonian Simulation & 5 \\
Initialization & 4 \\
Oracle & 4 \\
Basis Change & 3 \\
Circuit Construction Utility & 3 \\
\midrule
\textbf{Total} & \textbf{54} \\
\midrule
\multicolumn{2}{c}{\textbf{Qiskit}} \\
\midrule
Quantum Arithmetic & 26 \\
Circuit Construction Utility & 26 \\
Variational Quantum Algorithm (VQA) & 7 \\
Oracle & 5 \\
Quantum Logical Operators & 5 \\
Initialization & 4 \\
Basis Change & 3 \\
Data Encoding & 3 \\
Hamiltonian Simulation & 2 \\
Grover & 1 \\
\midrule
\textbf{Total} & \textbf{82} \\
\bottomrule
\end{tabularx}
\end{table}

The top 10 patterns for each framework are summarized in Table \ref{tab:combined_quantum_patterns}. The complete list can be found in the supplementary material.
During the classification process, we could identify patterns not previously documented in the literature. These patterns are detailed in the next Section.

\subsection{New Quantum Computing patterns}
\label{new_quantum_patterns}
When investigating the reference list of patterns (Section \ref{sub:pattern_atlas}), we identified that it has five specific data encoding methods as individual patterns (Angle Encoding, Amplitude Encoding, Base Encoding, Matrix Encoding and QRAM Encoding). However, this classification does not capture the full breadth of the field, as the literature on data encoding for QML describes a much larger variety of techniques \cite{math12213318, RathEncoding2024}. To address this, we introduced a more general pattern called \textbf{Data Encoding}, defined as any method designed to map classical information onto a quantum state.

Similarly, we created the unified \textbf{Basis Change} pattern after observing how certain operations with similar nature were categorized differently. For instance, the Quantum Fourier Transform (QFT) is classified as a standalone pattern \cite{Stiliadou2025_FundamentalPatterns}, while the Hadamard transform is typically considered part of the ``Uniform Superposition'' pattern \cite{LeymannQuantumAlgorithms}, even though both perform a basis change. Our analysis also identified uncategorized operations in Classiq, such as the quantum discrete cosine (DCT) and sine (DST) transforms. These are quantum versions of the classical Discrete Cosine and Sine Transforms, which change a state's basis using real-valued sinusoidal functions for applications in signal and image processing \cite{klappenecker2001discretecosinetransformsquantum}. Since QFT, Hadamard, DCT, and DST all share the core purpose of transforming between bases, we grouped them under the ``Basis Change'' pattern. This provides a more consistent and functionally descriptive classification.

Continuing the investigation by looking in the concepts extracted from the sourcing frameworks and comparing them to the reference list of patterns, we observed that several components could not be directly mapped to an existing pattern. We then defined new patterns, more specific to the nature of the concept being classified. These new patterns are described below.

\begin{enumerate}
    \item \textbf{Quantum Amplitude Estimation (QAE)}. This is an algorithm to estimate the probability of a specific outcome in a quantum state \cite{QAEBrassard_2002}. 
    
    \item \textbf{Linear Combination of Unitaries (LCU)}. A technique to construct a unitary operation by expressing it as a weighted sum of simpler, known unitaries. It uses an ancillary register to control the application of each component unitary \cite{LCU2012}.
    
    \item \textbf{Quantum Arithmetic}. We created this pattern to categorize quantum concepts related to classical arithmetic operations such as addition and multiplication on numbers encoded in the basis states of quantum registers, enabling computation on a superposition of values.
    
    \item \textbf{Quantum Logical Operators}. This pattern categorizes components that perform classical Boolean logic operations on quantum states. It includes operations such as logical AND, OR, XOR, and the inner product of quantum registers. These operators serve as building blocks for constructing more complex quantum algorithms, particularly in the implementation of oracles.
    
    \item \textbf{Circuit Construction Utility}. This pattern is about routines and techniques to create and manipulate quantum circuits such as swapping qubits or applying a gate to multiple wires, rather than implementing a full algorithm.
    
    \item \textbf{Domain Specific Application}. We defined this pattern to classify high-level applications or subroutines designed to solve specific problems within a particular domain. These components are typically composite, orchestrating several more quantum algorithms and patterns to provide an end-to-end solution.
    
    \item \textbf{Hamiltonian Simulation}. This pattern is about the simulation of the time evolution of a quantum system governed by a Hamiltonian $H$, with the goal of implementing the unitary operator $U(t) = e^{-iHt}$. In our analysis, instances of this pattern were identified through implementations of approximation methods such as the 
    Trotter-Suzuki decomposition \cite{Berry_2006}.
\end{enumerate}


In summary, this process of aggregation and addition of new patterns has updated the original pattern catalog. Starting from 59 patterns, we unified 5 existing patterns into the Data Encoding pattern, while 2 existing patterns were unified into the Basis Change category. We then introduced the 7 new patterns detailed in this section. As a result, our refined and expanded catalog now contains a total of 61 patterns.

\subsection{GitHub search for QC projects and repositories}
\label{sub:github_search}
This section describes the process we followed to select the QC projects analyzed in our work.
The process to select these projects is based on a Python script that uses the GitHub API to identify, filter and select the target repositories. 

The selection process began by querying the GitHub API for repositories that matched a set of specific search terms and the result of each query was joined in a single set of repositories. The specific search terms were: "\textit{topic: quantum-computing language:Python}", "\textit{topic: quantum-machine-learning language:Python}", and "\textit{topic: quan\-tum-algorithms language:Python}".

To prioritize projects, the results from each query were sorted by the number of stars in descending order. 
Subsequently, the candidate repositories were filtered based on the following criteria:

\noindent  \textbf{(1) Maturity and community engagement.} We considered projects with a minimum of \textbf{30 stars} and \textbf{10 contributors} to avoid small projects with low impact.
        
\noindent \textbf{(2) Project status and activity.} Repositories had to be actively maintained, non-archived, and must have received a code push within the preceding \textbf{12 months}. This filter excluded some legacy repositories, such as \texttt{rigetti/grove} (last updated in late 2021). However, we manually reinstated this specific repository for further analysis because it contains implementations and tests for many textbook quantum algorithms.        

\noindent \textbf{(3) Relevance for this study.} We removed repositories related to books and learning resources for QC.

The final dataset contains 80 projects whose main metrics and codebase statistics are detailed in the supplemental material.
The statistics for the Cirq ecosystem (including cirq-core, cirq-google, etc.) are presented as a single entry, as they are developed within the quantumlib/Cirq monorepo (single repository with multiple subfolders representing individual modules). Therefore, metrics such as stars and contributors reflect the activity and popularity of the entire Cirq project.

\subsection{QC pattern matching in GitHub repositories}
\label{QCPatternMatchingGithub}

This stage identifies the QC concepts and patterns from our reference list obtained as described in Section~\ref{sub:identifyQuantumConcepts} within the 80 open-source projects identified in Section \ref{sub:github_search}. 
The process involved three main phases:

\begin{enumerate}
    \item \textbf{Preparation of the knowledge base}

    First, we created a knowledge base of the QC concepts previously extracted from the reference frameworks (PennyLane, Qiskit, and Classiq SDK, as described in Section \ref{sub:identifyQuantumConcepts}). Using the \texttt{all-mpnet-base-v2} Sentence Transformer model \cite{allmpnetbasev2}, we generated two sets of numerical vector embeddings for each concept:
    \begin{itemize}
        \item One set of embeddings for the \textbf{concept names} (e.g., the function or class name).
        \item A second set of embeddings for the \textbf{concept summaries} (i.e., their docstrings).
    \end{itemize}
    This gave us two distinct vector databases to represent the meaning of our reference concepts: one for code-level syntax and one for natural language descriptions. Our choice of a local embedding model over a Large Language Model (LLM) was motivated by the goal of replicability. The model can be downloaded and executed on standard CPUs, which avoids relying on Graphical Processing Units (GPU) and the costs associated with API-based LLMs. Moreover, the embedding model is deterministic and not generative, so the results are not affected by the random output variations common in generative LLMs \cite{LLMHallucination2025}.

    \item \textbf{Preparation of the Target Code}

    Our analysis focused on finding QC concepts in practical applications rather than in framework-level code. Since Jupyter Notebooks are commonly used to demonstrate framework features by solving practical problems, we targeted them specifically.
    \begin{itemize}
        \item We searched the 80 target projects and found \textbf{985 Jupyter Notebooks (\texttt{.ipynb} files)}.
        \item These notebooks were then converted into standard Python scripts (\texttt{.py} files) to make them accessible to our code parser, preserving their original folder structure in a separate output directory.
    \end{itemize}

    \item \textbf{Semantic search and matching}

    With both the knowledge base and the target scripts prepared, we performed two parallel semantic searches on each Python script:
    \begin{itemize}
        \item \textbf{Matching by code elements:} We parsed each script's source code using Python's Abstract Syntax Tree (\texttt{ast}) module to extract all function and method calls. We generated embeddings for these call names and compared them against our concept name embeddings using cosine similarity. A match was recorded if the similarity score exceeded a threshold of 0.95.

        \item \textbf{Matching by comments:} We extracted all comments from each script and consolidated them into a single text block. We then generated an embedding for this block and compared it against our concept summary embeddings. By creating one embedding for the entire block of comments, the script aims to capture the overall topic or purpose of the entire converted notebook (Python file), which is much more likely to have a meaningful similarity to the summary of a quantum concept. A match was recorded if the cosine similarity score was above a lower threshold of \textbf{0.7}.
    \end{itemize}

These threshold values were chosen after experimentation, as they provided the best balance between minimizing false positives and avoiding an overly restrictive search. The final output is a CSV file detailing every detected match.
One should note that the search directly identifies low-level quantum concepts from our reference list (as defined in Section \ref{sub:identifyQuantumConcepts}) within the code of the target projects. The corresponding high-level pattern, which was assigned manually (as explained in Section \ref{sub:pattern_matching}), is then included as associated metadata. Each record in this file contains the file path, the matched QC concept, its pattern, the type of match (name or summary), the specific text that triggered the match, and the similarity score.

Following the semantic search, the final step in our methodology was to process the raw match data to generate a summary that addresses our research questions. We used a dedicated analysis script to ingest the raw output file (\textit{quantum\_concept\_matches\_with\_patterns.csv}) and transform the individual data points into meaningful statistics.

The script's primary function was to aggregate the matches and calculate key metrics. The main analyses performed were:

\begin{itemize}
    \item \textbf{Overall summary:} It calculated high-level statistics, including the total number of QC concepts detected, the number of unique projects in which concepts were found, and the performance of our two match types (`name` vs. `summary`).

    \item \textbf{Source and adoption analysis:} The script analyzed the origin and usage of concepts. It quantified which of the source frameworks (Qiskit, PennyLane, or Classiq) the matched concepts came from and which of the 80 target projects they were found in.

    \item \textbf{Pattern prevalence:} It determined the prevalence of each high-level QC pattern by counting its occurrences. This allowed us to identify which architectural and design patterns are most frequently used across the open-source ecosystem.

    \item \textbf{Gap analysis:} Finally, it performed a gap analysis by identifying which patterns, although defined in the source frameworks, were never detected in any of the target projects. This shows possible discrepancies between the patterns available in libraries and those adopted in practice.
\end{itemize}

The outcome of this script is a structured text report (\textit{final\_pattern\_report.txt}) that consolidates all these findings, providing the data for our final discussion and conclusions.

\end{enumerate}

\section{Results and discussion}
\label{sec:results}

This section presents and discusses the findings of our mining process, beginning with an analysis of our knowledge base and then moving to the results from the target projects.

First, we analyzed the patterns required to classify the concepts within our three source frameworks. From our extended catalog of 61 patterns, a total of 25 patterns were needed. The distribution of these patterns across the frameworks varied, with 19 patterns identified in Classiq, 15 in PennyLane, and 11 in Qiskit. From this subset, 9 patterns were common to all three frameworks. 
Next, our semantic search tool analyzed the 80 target projects, identifying 573 matches for quantum concepts across 251 unique files. These matches corresponded to 117 concepts from our knowledge base.

A total of 22 of the 24 patterns identified in the source frameworks were also found in the target projects. The two patterns found only in the frameworks but not in the practitioner code were \textbf{Function Table}, and \textbf{Schmidt Decomposition}.
The 10 most frequently matched patterns are listed in Table \ref{tab:patterns-by-match-count} and the complete list is in the supplementary material.

\begin{table}[ht]
\centering
\caption{Frequency of Quantum Patterns by match count}
\label{tab:patterns-by-match-count}
\begin{tabular*}{\columnwidth}{{@{} l @{\extracolsep{\fill}} r @{}}}
\toprule
\textbf{Pattern} & \textbf{Total Matches} \\
\midrule
Basis Change & 161 \\
Quantum Phase Estimation (QPE) & 57 \\
Domain Specific Application & 53 \\
Quantum Arithmetic & 43 \\
Circuit Construction Utility & 40 \\
Data Encoding & 37 \\
Amplitude Amplification & 26 \\
Initialization & 26 \\
Hamiltonian Simulation & 20 \\
Variational Quantum Algorithm (VQA) & 19 \\
\bottomrule
\end{tabular*}
\end{table}

For the similarity search, the matches were categorized into two types:
\begin{itemize}
\item \textbf{Name matches:} 384 matches were found by comparing code elements (e.g., function calls) to concept names. These matches had a high average similarity score of 0.99, indicating direct use of library components.
\item \textbf{Summary matches:} 189 matches were found by comparing code comments to concept summaries. These had an average similarity score of 0.70, suggesting discussion or explanation of a concept.
\end{itemize}

The results of our mining process provide a snapshot of how high-level quantum computing patterns are used in open-source software. In what follows, we interpret the key findings, answer our proposed research questions, consider their implications for the quantum computing community, and acknowledge the limitations of our study.

\vspace*{0.15cm}

\noindent \textit{RQ1. What are the most common QC concepts used by practitioners in popular open-source frameworks?}

\label{sub:RQ1}
The most common quantum computing abstractions are the basic algorithmic building blocks. As shown in Table \ref{tab:top-quantum-concepts}, concepts related to the Quantum Fourier Transform (QFT), Quantum Phase Estimation (QPE), and the Hadamard Transform were the most frequently detected. Implementations of these concepts from all three source frameworks (Classiq, Qiskit, and PennyLane) were found, which indicates they are common routines used across the quatum computing ecosystem.
For example, the \textit{hadamard\_transform} from Classiq was the single most detected concept. Similarly, QFT implementations from Qiskit, PennyLane, and Classiq appeared in the top results.

One of the differences of the three analyzed frameworks is the Classiq's emphasis on high-level, reusable functions. Two of its most frequent concepts, \textit{hadamard\_transform} and \textit{apply\_to\_all}, are examples of such functional paradigm. 
In contrast, PennyLane and Qiskit, for instance, use a gate-based approach to implement the same concepts.
From the three frameworks used to source the concepts, there is a clear predominance of Classiq on the list, with 11 of the 20 top concepts being identified in the target projects\footnote{File \url{https://anonymous.4open.science/r/quantum_patterns/docs/pdfs/final_pattern_report.pdf}, \textbf{Table VI -- Top matched concepts} of the supplementary material}. The focus on higher-level abstractions as described above, the extensive library of algorithms implemented and available on GitHub (for instance, 19 of the algorithms listed on the Quantum Zoo website have implementations on Classiq), and the high quality of their code documentation are likely the reasons for such performance. These higher-level abstractions can be seen as a factor that lowers the entry barrier for practitioners without a physics background to enter the domain of quantum computing.
Documentation is an important factor, as we use it to find matches in the target projects.
As our methodology uses \textit{docstrings} to identify semantic matches, the quality of a framework's documentation impacts its visibility in this analysis.

The other high-level concepts listed in Table \ref{tab:top-quantum-concepts} are algorithmic components used to construct larger programs. For example, Quantum Phase Estimation (QPE) and the Quantum Fourier Transform (QFT) are subroutines within Shor's algorithm. QPE is used to solve the period-finding problem, and its implementation requires the QFT. Similarly, \textit{grover\_operator}, \textit{amplitude\_amplification}, and \textit{phase\_oracle} are building blocks for quantum search algorithms. Amplitude amplification is the general method, implemented by the iterative \textit{grover\_operator}, which uses a \textit{phase\_oracle} to identify the target state. Finally, functions like Qiskit's \textit{z\_feature\_map} and \textit{zz\_feature\_map} support QML by providing methods to encode classical data into quantum states.

\begin{table}[ht]
\centering
\caption{Top 20 most frequently matched Quantum Concepts}
\label{tab:top-quantum-concepts}
\begin{tabular*}{\columnwidth}{{@{} l l @{\extracolsep{\fill}} r @{}}}
\toprule
\textbf{Framework} & \textbf{Concept} & \textbf{Matches} \\
\midrule
Classiq & \texttt{...hadamard\_transform} & 52 \\
Classiq & \texttt{...qpe} & 30 \\
Classiq & \texttt{...qft} & 26 \\
Classiq & \texttt{...qpe\_flexible} & 26 \\
Qiskit & \texttt{...QFT} & 24 \\
Pennylane & \texttt{...QFT} & 21 \\
Classiq & \texttt{...qsvt} & 19 \\
Classiq & \texttt{...apply\_to\_all} & 16 \\
Classiq & \texttt{...suzuki\_trotter} & 15 \\
Qiskit & \texttt{...AND} & 11 \\
\bottomrule
\end{tabular*}
\end{table}

\begin{researchbox}
\textbf{Answer RQ1:} The most common abstractions are fundamental algorithmic building blocks. Concepts related to the Hadamard Transform, Quantum Phase Estimation (QPE), and the Quantum Fourier Transform (QFT) are the most frequently detected. Frameworks also provide these abstractions differently: Classiq's high-level, functional concepts are the most prominent, while frameworks like PennyLane offer specialized components for domains such as Quantum Machine Learning, and Qiskit focuses on Gate-based components.
\end{researchbox}


\noindent \textit{RQ2. What QC patterns appear in practice that are not yet documented in the literature?}
\label{sub:RQ2}

We defined 9 new patterns in Section \ref{sub:pattern_matching}. 
\begin{researchbox}

\textbf{Answer RQ2:} The new QC patterns defined in our work are: Basis Change, Circuit Construction Utility, Data Encoding, Domain Specific Application, Hamiltonian Simulation, Linear Combination of Unitaries, Quantum Amplitude Estimation, Quantum Arithmetic, and Quantum Logical Operators.
\end{researchbox}

\vspace*{0.15cm}

\noindent \textit{RQ3. What QC architectural patterns are observed in these frameworks?}
\label{sub:RQ3}

Our analysis of the sourcing frameworks (Section \ref{sub:identifyQuantumConcepts}) and further investigation of the patterns in the target projects allowed us to identify certain architectural patterns.
For instance, we observed that the workflows of classical Machine Learning (ML) have been adopted in the quantum domain. An example is the Qiskit Machine Learning library, which uses the API design of Scikit-learn, a popular classical ML framework \cite{PedrosaScikitLearn2011}. This design is based on the principle of Separation of Concerns to create a modular architecture for ML components, as documented by \citeauthor{SEDesignPatternsML2022} \cite{SEDesignPatternsML2022} and exemplified in Listing \ref{lst:qiskit_pattern_example}.
The idea is to provide high-level abstractions through a set of common interfaces, making the technology more accessible to non-specialists. Besides promoting modularity, this approach also makes the interoperability of quantum workflows with classical ML frameworks easier.
Furthermore, we observe the application of certain classical software design patterns. For instance, the optimizer (defined on line 3 and used on 13) object is an example of the \textbf{Strategy} Pattern \cite{DesignPatternsGof}, which allows different optimization algorithms to be selected and swapped at runtime. Another example is the high-level classifier class (\textit{EstimatorQNN} on line 10), which can be seen as an implementation of the classical \textbf{Façade} design pattern \cite{DesignPatternsGof} or its quantum counterpart, the \textbf{Classical-Quantum Interface} pattern \cite{Buehler2023_QuantumSoftwareEngineeringPatterns}. This class consists of a simplified interface that hides the underlying complexity of the quantum neural network and its training loop from the end-user.

We also observed the \textit{Quantum Module} and \textbf{Quantum Module Template} patterns \cite{Buehler2023_QuantumSoftwareEngineeringPatterns} in the three sourcing frameworks, which grouped quantum-related concepts into the same folder (sometimes called templates, as the case of PennyLane) and allowed us to identify and extract quantum related concepts from these frameworks automatically. These frameworks also promote segregation of responsibility implementing the pattern \textbf{Quantum-Classic Split} \cite{LeymannQuantumAlgorithms}. In this pattern, classical parts of certain algorithms are kept in the classical framework code, while the quantum parts are shifted to modules and circuits that are executed in a Quantum Computer. Consequently, the \textbf{Hybrid Module} pattern is also present to all three frameworks, as their primary function is to orchestrate the interaction between classical code and quantum processing units (real or simulated). We also observed an emphasis on quality assurance, with the \textbf{Quantum Application Testing} pattern manifested through test suites at different levels (e.g., unit, integration) in all frameworks. Finally, it is important to distinguish these architectural patterns from application domains. Although we found concepts related to the \textbf{Quantum Classification} pattern \cite{Stiliadou2025_QMLPatterns}, we categorize this as a specific task type rather than a pattern.

\begin{lstlisting}[
    language=Python,
    caption={
        An extract from the file \url{test\_neural\_network\_classifier.py} from the Qiskit Machine Learning framework.
    },
    label={lst:qiskit_pattern_example}
]
def test_classifier_with_estimator_qnn(self, opt, loss, cb_flag):
    ...
    optimizer = self._create_optimizer(opt)
    ...
    feature_map = zz_feature_map(num_inputs)
    ansatz = real_amplitudes(num_inputs, reps=1)
    qc = QuantumCircuit(num_inputs)
    qc.compose(feature_map, inplace=True)
    qc.compose(ansatz, inplace=True)
    qnn = EstimatorQNN(
        circuit=qc, input_params=feature_map.parameters, weight_params=ansatz.parameters
    )
    classifier = self._create_classifier(qnn, ansatz.num_parameters, optimizer, loss, callback)
    ...
    classifier.fit(X, y)
    ...
\end{lstlisting}

\begin{researchbox}
\textbf{Answer RQ3:} Our analysis reveals an adoption of architectural patterns from classical software engineering, particularly from the machine learning domain. This includes high-level designs, like Scikit-learn's modular API, and established design patterns such as Strategy and Façade. Alongside these, we identified core quantum-specific patterns like Quantum Module, Quantum-Classic Split, and Hybrid Module.
\end{researchbox}

\vspace*{0.05cm}

\noindent \textit{RQ4. How are QC patterns applied to develop practical applications, as opposed to internal framework code?}
\label{sub:RQ4}

By analyzing Jupyter Notebooks, we focused on how patterns are used in practical examples rather than internal framework code. The results show that developers use patterns in two main ways.
First, the high number of direct name matches (384 out of 573) indicates that developers frequently import and use pre-built, high-level components from libraries like Qiskit, PennyLane, and Classiq to construct their programs. This suggests a move away from building algorithms from scratch with low-level gate operations.

Second, the most frequently applied patterns correspond to basic algorithmic building blocks and structural utilities. For instance, \textbf{Basis Change} (161 matches), \textbf{Quantum Phase Estimation} (QPE) (57 matches), and \textbf{Quantum Arithmetic} (43 matches) were the most common. This points to a common need for these core routines when constructing more complex quantum programs.
There were also patterns that could not be detected in any of the target projects: \textbf{Function Table}, and \textbf{Schmidt Decomposition}.

\begin{researchbox}

\textbf{Answer RQ4:} Our analysis shows that the most common patterns are basic building blocks such as Basis Change, Quantum Phase Estimation (QPE), and Quantum Arithmetic. We also observed framework specializations: PennyLane is prominent in Quantum Machine Learning patterns (e.g., QNN, VQA), while Qiskit contains more low-level circuit and logic patterns (e.g., Quantum Logical Operators).
\end{researchbox}

\subsection{A Hierarchy of abstractions in practice}
Our findings suggest that developers work with quantum patterns at three distinct levels of abstraction:

\noindent \textbf{Foundational layer}. At the lowest level, we observed frequent use of circuit construction utilities and basic operations. The high number of matches for concepts like hadamard\_transform and control-flow constructs like apply\_to\_all indicates that developers still engage in the detailed construction of quantum circuits, requiring fine-grained control.

\noindent \textbf{Algorithmic primitives layer.} The middle layer consists of reusable building blocks that implement well-known algorithms. The prevalence of concepts like Quantum Phase Estimation (QPE) and the Quantum Fourier Transform (QFT) shows that developers rely on these standard subroutines rather than re-implementing them. This layer also includes patterns for data encoding, such as those used in Quantum Machine Learning.

\noindent \textbf{Application layer.} At the highest level, we found patterns that solve domain-specific problems. The detection of concepts like `QuantumMonteCarlo` (finance) and `QAOAEmbedding` (optimization) suggests a growing maturity in the ecosystem, where developers are using quantum software to address specific, real-world problems.

This hierarchy indicates that the community is beginning to move from low-level circuit manipulation towards solving problems at higher levels of abstraction, which is a positive sign for the usability of quantum software.

When categorizing each quantum concept into the different patterns, we observed the need to re-evaluate patterns previously identified as ``QML Patterns.'' Terms like ``Quantum Clustering'' \cite{Stiliadou2025_QMLPatterns}, ``Quantum Kernel Estimation (QKE)'' \cite{PatternsHybridQML2021}, and ``Quantum Neural Networks (QNN)'' \cite{Stiliadou2025_QMLPatterns} are better described as quantum implementations of high-level machine learning tasks, rather than as distinct quantum computing patterns. A pattern provides a reusable blueprint for a software architecture, whereas a task defines a problem category. For example, ``Quantum Classification'' \cite{Stiliadou2025_QMLPatterns} does not represent a single, repeatable design. Instead, classification is a broad objective that can be addressed by a diverse set of models (such as a Quantum Support Vector Machine or a Variational Quantum Classifier), each with its own distinct implementation and underlying architectural patterns. 

Nevertheless, we decided to consider these task-level concepts in our analysis because QML frameworks such as PennyLane present them as high-level reusable components. This suggests a trend toward providing developers with abstractions beyond the traditional circuit-based model. In our classification process (described in Section \ref{sub:pattern_matching}), QNN was the only task-related pattern that we identified. Our subsequent analysis (described in Section \ref{QCPatternMatchingGithub}) shows that QNN was present in five of the analyzed projects: Cirq, classiq-library, qiskit-machine-learning, quimb, and tensorcircuit. 

\subsection{Implications of the Findings}

Our results have several implications. For framework developers, the widespread adoption of circuit construction utilities suggests that continued investment in tools that simplify circuit building and manipulation is valuable. We also observed a possible specialization of frameworks, as for instance, in the case of PennyLane with its focus in QML and the use of QML patterns by the community.

For practitioners and researchers, our findings show that a core set of primitives and modules are used to build different types of quantum applications. For instance, when investigating the matches found in the target projects, we observe occurrences of the patterns in financial applications, natural sciences, optimization, vehicle routing, and reinforcement learning.

The observation that two patterns were not used at all might indicate a gap between the patterns offered by libraries and the actual needs of developers, which could guide future research.

One of the contributions of this work is its focus on practical applications. By analyzing Jupyter Notebooks instead of internal framework code, we provide a view of how quantum patterns are used to solve real-world problems. This analysis resulted in an annotated database that maps high-level patterns to concrete implementations, serving as a reference for researchers, practitioners and students. Furthermore, we extend the existing pattern catalog by identifying and documenting nine new patterns. 
Finally, the semantic search tool developed for this research is reusable and can be applied to future studies to track the evolution of pattern usage or analyze new quantum software repositories.

\subsection{Threats to validity}
\label{sec:threats}
This section summarizes the choices and trade-offs made during our investigation and their potential impact on our conclusions.

\noindent  \textbf{Internal Validity.} Our knowledge base is derived from three frameworks: Qiskit, PennyLane, and Classiq. While this is not an exhaustive sample, they were chosen to cover distinct design paradigms (object-oriented, functional, and model-based synthesis). This architectural diversity provided different implementations of the same concepts, which enriched the dataset for our semantic search tool. The manual categorization of patterns also introduced a risk of bias. We mitigated this by having three authors perform the task independently before discussing the results to reach a consensus. Finally, we set high matching thresholds in our semantic search to minimize false positives. This choice proved effective, identifying 573 concepts with high average scores for both names (0.99) and summaries (0.69), though it may have missed some less common patterns.

\noindent \textbf{External Validity.} Our study is limited to Python-based frameworks, which might overlook patterns specific to other languages like C++, Q\#, and Julia. To mitigate this, our analysis was not limited to mainstream libraries; we examined a broad range of 80 diverse projects to have a more representative sample. Extending this analysis to other languages remains a direction for future work.

\section{Related Work}
\label{related_work}

Recent studies have also investigated how quantum computing patterns are used in open-source projects. For example, \citeauthor{fernandez2025exploring} \cite{fernandez2025exploring} conducted a  case study on 2,610 Qiskit programs from GitHub. Their work focused specifically on detecting four foundational patterns: Initialization, Superposition, Entanglement, and Oracle—using. The authors used the QCPD tool developed by \citeauthor{RomeroAutomata} \cite{RomeroAutomata}. 

Working on another approach aimed to investigate the manifestation of quantum software patterns in open-source projects, \citeauthor{CastilloPreliminaryStudy2024} \cite{CastilloPreliminaryStudy2024} performed a code repository analysis. They examined 80 source files from Qiskit and OpenQASM, searching for five specific patterns: Initialization, Uniform Superposition, Oracle, Entanglement, and Uncompute. The study found that Initialization and Uniform Superposition were the most frequently implemented patterns in the analyzed codebases. Similarly, \citeauthor{shen2025quantumpatterndetectionaccurate} \cite{shen2025quantumpatterndetectionaccurate} created a tool that can identify 8 quantum patterns (Uniform Superposition, Creating Entanglement, Basis Encoding, Angle Encoding, Amplitude Encoding, Quantum Phase Estimation, Uncompute, Post Selective Measurement) using static and dynamic analysis. The authors study algorithms from Qiskit 0.45.0  and MQT Bench \cite{Quetschlich_2023}. The list of algorithms analyzed is limited to 20 elements (Adder w. overflow, Adder w/o overflow, Amplitude Encoding, Amplitude Estimation, Deutsch-Jozsa, GHZ, Graph State, Grover, HHL, Multiplier, QAOA, QFT, QFT w. entanglement, OPE, Quantum Walk, Real Amplitudes, Shor, SU2 Ansatz, VQE, and W-State).

In terms of Quantum Software Architecture patterns \citeauthor{aktar2025decisionmodelsselectingarchitecture} \cite{aktar2025decisionmodelsselectingarchitecture} propose a framework to support the selection of architectural patterns in quantum software that uses six decision models. These models are based on six quantum software architecture patterns defined by the authors: Communication, Decomposition, Data Processing, Fault Tolerance, Integration and Optimization, and Algorithm Implementation. 

Our work extends previous research through four contributions. First, we broaden the scope of analysis, examining 80 projects, whereas prior studies focused on only a few frameworks. Second, our methodology is practitioner-focused, as we analyze 985 Jupyter Notebooks to understand how patterns are used in real-world applications, not just within the frameworks themselves. Third, to enable this research, we developed a semantic search tool to identify patterns in open-source Python projects. Finally, this practical approach resulted in the discovery and documentation of 9 new patterns, expanding the known pattern catalog.

\section{Conclusions and future work}
\label{sec:conclusions}

Quantum Computing has evolved from a theoretical field into a practical domain active in both academia and industry. In this work, we analyzed how practitioners use patterns for developing quantum applications by mapping documented patterns to three open-source frameworks (Classiq, Qiskit, and PennyLane) and using the resulting concepts to analyze QC-related GitHub projects.
One of our contributions is a catalog of quantum computing concepts categorized into patterns that can support researchers in building higher-level applications and serve as an educational tool for students to connect theory with real-world code. Our categorization process also led to the definition of nine new patterns observed in the studied frameworks. The semantic tool developed as part of this work is a reusable artifact for finding quantum pattern implementations in other Python-based projects. 
Finally, we discuss insights on how QC patterns occur in 985 Jupyter Notebooks from 80 open-source projects.

This work can be extended in several key directions.
First, since our analysis currently focuses only on Python, expanding our tool to support other quantum programming languages such as Q\# and Julia is a natural next step. Additionally, the pattern categorization process is currently manual. A valuable improvement would be to develop a tool that automates this task, possibly using embedding models or LLMs. Finally, our semantic search component uses a high similarity threshold to reduce incorrect matches. A promising research path involves fine-tuning this component to lower the threshold, allowing it to find more relevant matches while keeping the number of false positives low.

\bibliographystyle{ACM-Reference-Format}
\bibliography{references}

@book{Hidary2019Quantum,
	author = {Hidary, Jack D.},
	publisher = {Springer},
	title = {Quantum {Computing}: An {Applied} {Approach}},
	year = {2019}}

@misc{qiskit2024,
      title={Quantum computing with {Q}iskit},
      author={Javadi-Abhari, Ali and Treinish, Matthew and Krsulich, Kevin and Wood, Christopher J. and Lishman, Jake and Gacon, Julien and Martiel, Simon and Nation, Paul D. and Bishop, Lev S. and Cross, Andrew W. and Johnson, Blake R. and Gambetta, Jay M.},
      year={2024},
      doi={10.48550/arXiv.2405.08810},
      eprint={2405.08810},
      archivePrefix={arXiv},
      primaryClass={quant-ph}
}

@misc{QuantumZoo,
  author       = {Jordan, Stephen},
  title        = {The Quantum Algorithm Zoo},
  howpublished = {\url{http://quantumalgorithmzoo.org}},
  year         = {2024},
  note         = {Accessed: June 2025}
}

@misc{classiq-library2024,
      title={Design and synthesis of scalable quantum programs},
      author={Goldfriend, Tomer and Reichental, Israel and Naveh, Amir and Gazit, Lior and Yoran, Nadav and Alon, Ravid and Ur, Shmuel and Lahav, Shahak and Cornfeld, Eyal and Elazari, Avi and Emanuel, Peleg and Harpaz, Dor and Michaeli, Tal and Erez, Nati and Preminger, Lior and Shapira, Roman and Garcell, Erik Michael and Samimi, Or and Kisch, Sara and Hallel, Gil and Kishony, Gilad and Wingerden, Vincent van and Rosenbloom, Nathaniel A. and Opher, Ori and Vax, Matan and Smoler, Ariel and Danzig, Tamuz and Schirman, Eden and Sella, Guy and Cohen, Ron and Garfunkel, Roi and Cohn, Tali and Rosemarin, Hanan and Hass, Ron and Jankiewicz, Klem and Gharra, Karam and Roth, Ori and Azar, Barak and Asban, Shahaf and Linkov, Natalia and Segman, Dror and Sahar, Ohad and Davidson, Niv and Minerbi, Nir and Naveh, Yehuda},
      year={2024},
      doi={10.48550/arXiv.2412.07372},
      eprint={2412.07372},
      archivePrefix={arXiv},
      primaryClass={quant-ph}
}

@misc{pennylane_qml_github,
  author       = {{The PennyLane team and contributors}},
  title        = {{PennyLane QML: Demonstrations, tutorials, and implementations of quantum machine learning}},
  howpublished = {\url{https://github.com/PennyLaneAI/qml}},
  publisher    = {GitHub},
  year         = {2025},
  note         = {Accessed: 2025-08-21}
}

@inproceedings{LeymannQuantumAlgorithms,
    author = {Leymann, Frank},
    title = {Towards a Pattern Language for Quantum Algorithms},
    booktitle = {Quantum Technology and Optimization Problems},
    year = {2019},
    pages = {218--230},
    doi = {10.1007/978-3-030-14082-3_19},
    series = {Lecture Notes in Computer Science (LNCS)},
    volume = {11413},
    publisher = {Springer International Publishing},
    address = {Cham}
}

@inproceedings{Buehler2023_QuantumSoftwareEngineeringPatterns,
author = {B{\"u}hler, Fabian and Barzen, Johanna and Beisel, Martin and
Georg, Daniel and Leymann, Frank and Wild, Karoline},
title = {{Patterns for Quantum Software Development}},
booktitle = {Proceedings of the 15th International Conference on Pervasive
Patterns and Applications (PATTERNS 2023)},
year = {2023},
month = jun,
pages = {30--39},
publisher = {Xpert Publishing Services (XPS)},
isbn = {978-1-68558-049-0}
}

@article{fernandez2025exploring,
  title={Exploring design patterns in quantum software: a case study},
  author={Fern{\'a}ndez-Osuna, Miriam and P{\'e}rez-Castillo, Ricardo and Cruz-Lemus, Jos{\'e} A. and Baczyk, Michal and Piattini, Mario},
  journal={Computing},
  volume={107},
  pages={111},
  year={2025},
  publisher={Springer},
  doi={10.1007/s00607-025-01467-2},
  url={https://doi.org/10.1007/s00607-025-01467-2}
}

@article{RomeroAutomata,
    author = {Romero, Francisco P. and Cruz-Lemus, Jos\'{e} A. and Jim\'{e}nez-Fern\'{a}ndez, Sergio and Piattini, Mario},
    title = {Automata-Based Quantum Circuit Design Patterns Identification: A Novel Approach and Experimental Verification},
    journal = {International Journal of Software Engineering and Knowledge Engineering},
    volume = {34},
    number = {09},
    pages = {1415-1439},
    year = {2024},
    doi = {10.1142/S0218194024410031},
    URL = {https://doi.org/10.1142/S0218194024410031},
    eprint = {https://doi.org/10.1142/S0218194024410031}
}

@misc{shen2025quantumpatterndetectionaccurate,
      title={Quantum Pattern Detection: Accurate State- and Circuit-based Analyses}, 
      author={Julian Shen and Joshua Ammermann and Christoph König and Ina Schaefer},
      year={2025},
      eprint={2501.15895},
      archivePrefix={arXiv},
      primaryClass={quant-ph},
      url={https://arxiv.org/abs/2501.15895}, 
}

@article{Quetschlich_2023,
   title={MQT Bench: Benchmarking Software and Design Automation Tools for Quantum Computing},
   volume={7},
   ISSN={2521-327X},
   url={http://dx.doi.org/10.22331/q-2023-07-20-1062},
   DOI={10.22331/q-2023-07-20-1062},
   journal={Quantum},
   publisher={Verein zur Forderung des Open Access Publizierens in den Quantenwissenschaften},
   author={Quetschlich, Nils and Burgholzer, Lukas and Wille, Robert},
   year={2023},
   month=jul, pages={1062} }

@inproceedings{Stiliadou2025_QMLPatterns,
 author = {Stiliadou, Lavinia and Barzen, Johanna and Beisel, Martin and
Leymann, Frank and Weder, Benjamin},
 title = {{Patterns for Quantum Machine Learning}},
 booktitle = {Proceedings of the 17\textsuperscript{th} International
Conference on Pervasive Patterns and Applications (PATTERNS
2025)},
 year = {2025},
 month = apr,
 pages = {7-14},
 publisher = {Xpert Publishing Services (XPS)},
 Isbn = {78-1-68558-263-0}
}

@misc{klymenko2024architecturalpatternsdesigningquantum,
      title={Architectural Patterns for Designing Quantum Artificial Intelligence Systems}, 
      author={Mykhailo Klymenko and Thong Hoang and Xiwei Xu and Zhenchang Xing and Muhammad Usman and Qinghua Lu and Liming Zhu},
      year={2024},
      eprint={2411.10487},
      archivePrefix={arXiv},
      primaryClass={cs.SE},
      url={https://arxiv.org/abs/2411.10487}, 
}

@INPROCEEDINGS{CastilloPreliminaryStudy2024,
  author={Pérez-Castillo, Ricardo and Fernández-Osuna, Miriam and Cruz-Lemus, José A. and Piattini, Mario},
  booktitle={2024 IEEE/ACM 5th International Workshop on Quantum Software Engineering (Q-SE)}, 
  title={A preliminary study of the usage of design patterns in quantum software}, 
  year={2024},
  volume={},
  number={},
  pages={41-48},
  doi={}
}

@online{PlanQK_QuantumPatterns_2024,
  author       = {{PlanQK}},
  title        = {Quantum Computing Patterns},
  year         = {2025},
  url          = {https://patternatlas.planqk.de/pattern-languages/af7780d5-1f97-4536-8da7-4194b093ab1d},
  urldate      = {2025-09-28}
}

@book{DesignPatternsGof,
    author = {Gamma, Erich and Helm, Richard and Johnson, Ralph and Vlissides, John},
    title = {Design patterns: elements of reusable object-oriented software},
    year = {1995},
    isbn = {0201633612},
    publisher = {Addison-Wesley Longman Publishing Co., Inc.},
    address = {USA}
}

@inproceedings{HarnessingPatterns2025,
    author = {Vietz, Daniel and Barzen, Johanna and Beisel, Martin and Leymann, Frank and Stiliadou, Lavinia and Weder, Benjamin},
    title = {Harnessing Patterns to Support the Development of Hybrid Quantum Applications},
    year = {2025},
    isbn = {9798400710124},
    publisher = {Association for Computing Machinery},
    address = {New York, NY, USA},
    url = {https://doi.org/10.1145/3731806.3731809},
    doi = {10.1145/3731806.3731809},
    booktitle = {Proceedings of the 2025 14th International Conference on Software and Computer Applications},
    pages = {40–46},
    numpages = {7},
    series = {ICSCA '25}
}

@inproceedings{PatternsQuantumErrorHandling2022,
    author = {Beisel, Martin and Barzen, Johanna and Leymann, Frank and
    Truger, Felix and Weder, Benjamin and Yussupov, Vladimir},
    title = {{Patterns for Quantum Error Handling}},
    booktitle = {Proceedings of the 14\textsuperscript{th} International
    Conference on Pervasive Patterns and Applications (PATTERNS
    2022)},
    year = {2022},
    month = apr,
    pages = {22-30},
    publisher = {Xpert Publishing Services (XPS)},
    Isbn = {978-1-61208-953-9}
}

@inproceedings{PatternsCircuitCutting2023,
    author = {Bechtold, Marvin and Barzen, Johanna and Beisel, Martin and Leymann, Frank and Weder, Benjamin},
    title = {Patterns for Quantum Circuit Cutting},
    year = {2023},
    isbn = {9781941652190},
    publisher = {The Hillside Group},
    address = {USA},
    booktitle = {Proceedings of the 30th Conference on Pattern Languages of Programs},
    articleno = {25},
    numpages = {12},
    location = {Monticello, IL, USA},
    series = {PLoP '23}
}

@inproceedings{Georg2023_PatternsQuantumExecution,
    Author = {Georg, Daniel and Barzen, Johanna and Beisel, Martin and Leymann, Frank and
    Obst, Julian and Vietz, Daniel and Weder, Benjamin and Yussupov, Vladimir},
    Title = {Execution Patterns for Quantum Applications},
    Booktitle = {Proceedings of the 18th International Conference on Software Technologies - ICSOFT},
    Year = {2023},
    Month = jul,
    Pages = {258--268},
    Publisher = {SciTePress},
    ISBN = {978-989-758-665-1},
    ISSN = {2184-2833},
    URL = {https://www.scitepress.org/Link.aspx?doi=10.5220%2f0012057700003538},
    DOI = {10.5220/0012057700003538},
    projects = {PlanQK,EniQmA,SeQuenC}
}

@misc{aktar2025decisionmodelsselectingarchitecture,
      title={Decision Models for Selecting Architecture Patterns and Strategies in Quantum Software Systems}, 
      author={Mst Shamima Aktar and Peng Liang and Muhammad Waseem and Amjed Tahir and Mojtaba Shahin and Muhammad Azeem Akbar and Arif Ali Khan and Aakash Ahmad and Musengamana Jean de Dieu and Ruiyin Li},
      year={2025},
      eprint={2507.11671},
      archivePrefix={arXiv},
      primaryClass={cs.SE},
      url={https://arxiv.org/abs/2507.11671}, 
}

@misc{QAEBrassard_2002,
   title={Quantum amplitude amplification and estimation},
   ISSN={0271-4132},
   url={http://dx.doi.org/10.1090/conm/305/05215},
   DOI={10.1090/conm/305/05215},
   journal={Quantum Computation and Information},
   publisher={American Mathematical Society},
   author={Brassard, Gilles and Høyer, Peter and Mosca, Michele and Tapp, Alain},
   year={2002},
   pages={53–74} }

@misc{klappenecker2001discretecosinetransformsquantum,
      title={Discrete Cosine Transforms on Quantum Computers}, 
      author={Andreas Klappenecker and Martin Roetteler},
      year={2001},
      eprint={quant-ph/0111038},
      archivePrefix={arXiv},
      primaryClass={quant-ph},
      url={https://arxiv.org/abs/quant-ph/0111038}, 
}

@article{LCU2012, 
    title={Hamiltonian Simulation Using Linear Combinations of Unitary Operations}, 
    volume={12},
   ISSN={1533-7146},
   url={http://dx.doi.org/10.26421/QIC12.11-12},
   DOI={10.26421/qic12.11-12},
   number={11 and 12},
   journal={Quantum Information and Computation},
   publisher={Rinton Press},
   year={2012},
   month=nov 
}

@article{MurilloQSE2025,
    author = {Murillo, Juan Manuel and Garcia-Alonso, Jose and Moguel, Enrique and Barzen, Johanna and Leymann, Frank and Ali, Shaukat and Yue, Tao and Arcaini, Paolo and P\'{e}rez-Castillo, Ricardo and Garc\'{\i}a-Rodr\'{\i}guez de Guzm\'{a}n, Ignacio and Piattini, Mario and Ruiz-Cort\'{e}s, Antonio and Brogi, Antonio and Zhao, Jianjun and Miranskyy, Andriy and Wimmer, Manuel},
    title = {Quantum Software Engineering: Roadmap and Challenges Ahead},
    year = {2025},
    issue_date = {June 2025},
    publisher = {Association for Computing Machinery},
    address = {New York, NY, USA},
    volume = {34},
    number = {5},
    issn = {1049-331X},
    url = {https://doi.org/10.1145/3712002},
    doi = {10.1145/3712002},
    month = may,
    articleno = {154},
    numpages = {48}
}

@article{PedrosaScikitLearn2011,
    author = {Pedregosa, Fabian and Varoquaux, Ga\"{e}l and Gramfort, Alexandre and Michel, Vincent and Thirion, Bertrand and Grisel, Olivier and Blondel, Mathieu and Prettenhofer, Peter and Weiss, Ron and Dubourg, Vincent and Vanderplas, Jake and Passos, Alexandre and Cournapeau, David and Brucher, Matthieu and Perrot, Matthieu and Duchesnay, \'{E}douard},
    title = {Scikit-learn: Machine Learning in Python},
    year = {2011},
    issue_date = {2/1/2011},
    publisher = {JMLR.org},
    volume = {12},
    number = {null},
    issn = {1532-4435},
    journal = {J. Mach. Learn. Res.},
    month = nov,
    pages = {2825–2830},
    numpages = {6}
}

@ARTICLE{SEDesignPatternsML2022,
    author={Washizaki, Hironori and Khomh, Foutse and Gueheneuc, Yann-Gael and Takeuchi, Hironori and Natori, Naotake and Doi, Takuo and Okuda, Satoshi},
    journal={ Computer },
    title={{ Software-Engineering Design Patterns for Machine Learning Applications }},
    year={2022},
    volume={55},
    number={03},
    ISSN={1558-0814},
    pages={30-39},
    doi={10.1109/MC.2021.3137227},
    url = {https://doi.ieeecomputersociety.org/10.1109/MC.2021.3137227},
    publisher={IEEE Computer Society},
    address={Los Alamitos, CA, USA},
    month=mar
}

@Article{math12213318,
    AUTHOR = {Ranga, Deepak and Rana, Aryan and Prajapat, Sunil and Kumar, Pankaj and Kumar, Kranti and Vasilakos, Athanasios V.},
    TITLE = {Quantum Machine Learning: Exploring the Role of Data Encoding Techniques, Challenges, and Future Directions},
    JOURNAL = {Mathematics},
    VOLUME = {12},
    YEAR = {2024},
    NUMBER = {21},
    ARTICLE-NUMBER = {3318},
    URL = {https://www.mdpi.com/2227-7390/12/21/3318},
    ISSN = {2227-7390}
}

@article{RathEncoding2024,
	author = {Rath, Minati and Date, Hema},
	date = {2024/10/25},
	doi = {10.1140/epjqt/s40507-024-00285-3},
	id = {Rath2024},
	isbn = {2196-0763},
	journal = {EPJ Quantum Technology},
	number = {1},
	pages = {72},
	title = {Quantum data encoding: a comparative analysis of classical-to-quantum mapping techniques and their impact on machine learning accuracy},
	url = {https://doi.org/10.1140/epjqt/s40507-024-00285-3},
	volume = {11},
	year = {2024}
}

@InProceedings{PatternsHybridQML2021,
    author="Weigold, Manuela
    and Barzen, Johanna
    and Leymann, Frank
    and Vietz, Daniel",
    editor="Barzen, Johanna",
    title="Patterns for Hybrid Quantum Algorithms",
    booktitle="Service-Oriented Computing",
    year="2021",
    publisher="Springer International Publishing",
    address="Cham",
    pages="34--51",
    isbn="978-3-030-87568-8"
}

@inproceedings{Stiliadou2025_FundamentalPatterns,
    Title = {{Fundamental Patterns for Composing Quantum Algorithms}},
    Author = {Stiliadou, Lavinia and Barzen, Johanna and Beisel, Martin and Leymann, Frank and Weder, Benjamin},
    Year = 2025,
    Month = jun,
    Booktitle = {Proceedings of the 1\textsuperscript{st} International Conference on Quantum Software (IQSOFT 2025)},
    Publisher = {SciTePress},
    Pages = {61--69},
    Doi = {10.5220/0013555600004525},
    url = {https://www.scitepress.org/PublicationsDetail.aspx?ID=WOl9XPv0NrQ=&t=1},
    projects = {EniQmA,SeQuenC}
}

@misc{allmpnetbasev2,
    author = {Sentence Transformers},
    title = {all-mpnet-base-v2},
    year = {2021},
    publisher = {HuggingFace},
    url = {https://huggingface.co/sentence-transformers/all-mpnet-base-v2}
}

@inproceedings{reimers-2019-sentence-bert,
  title = "Sentence-BERT: Sentence Embeddings using Siamese BERT-Networks",
  author = "Reimers, Nils and Gurevych, Iryna",
  booktitle = "Proceedings of the 2019 Conference on Empirical Methods in Natural Language Processing",
  month = "11",
  year = "2019",
  publisher = "Association for Computational Linguistics",
  url = "https://arxiv.org/abs/1908.10084",
}

@article{Berry_2006,
   title={Efficient Quantum Algorithms for Simulating Sparse Hamiltonians},
   volume={270},
   ISSN={1432-0916},
   url={http://dx.doi.org/10.1007/s00220-006-0150-x},
   DOI={10.1007/s00220-006-0150-x},
   number={2},
   journal={Communications in Mathematical Physics},
   publisher={Springer Science and Business Media LLC},
   author={Berry, Dominic W. and Ahokas, Graeme and Cleve, Richard and Sanders, Barry C.},
   year={2006},
   month=dec, pages={359–371} 
}

@article{LLMHallucination2025,
    author = {Huang, Lei and Yu, Weijiang and Ma, Weitao and Zhong, Weihong and Feng, Zhangyin and Wang, Haotian and Chen, Qianglong and Peng, Weihua and Feng, Xiaocheng and Qin, Bing and Liu, Ting},
    title = {A Survey on Hallucination in Large Language Models: Principles, Taxonomy, Challenges, and Open Questions},
    year = {2025},
    issue_date = {March 2025},
    publisher = {Association for Computing Machinery},
    address = {New York, NY, USA},
    volume = {43},
    number = {2},
    issn = {1046-8188},
    url = {https://doi.org/10.1145/3703155},
    doi = {10.1145/3703155},
    month = jan,
    articleno = {42},
    numpages = {55}
}

\end{document}